\documentclass[notitlepage,onecolumn]{article}%
\usepackage{amsmath}
\usepackage{amsfonts}
\usepackage{amssymb}
\usepackage{graphicx}%
\newtheorem{theorem}{Theorem}

\newtheorem{example}[theorem]{Example}

\begin{document}

\title{A general method for debiasing a Monte Carlo estimator }
\author{Don McLeish\\{\small Department of Statistics and Actuarial Science,}\\{\small University of Waterloo }\\{\small Tel.: +519-8884567 Ext. 35534}\\{\small dlmcleis@uwaterloo.ca}}
\date{}
\maketitle

\begin{abstract}
Consider a stochastic process $X_{n},$ $n=0,1,2,...$such that $EX_{n}%
\rightarrow x_{\infty}$ as $n\rightarrow\infty.$ \ The sequence $\{X_{n}\}$
may be a deterministic one, obtained by using a numerical integration scheme,
\ \ or obtained from Monte-Carlo methods involving an approximation to an
integral, or a Newton-Raphson iteration to approximate the root of an equation
but we will assume that we can sample from the distribution of $X_{1}%
,X_{2},...X_{m}$ \ for finite $m$. \ \ We propose a scheme for unbiased
estimation of the limiting value $x_{\infty}$, \ together with estimates of
standard error and apply this to examples including numerical integrals,
root-finding and option pricing in a Heston Stochastic Volatility
model.\newline\textbf{Keywords and phrases: }Monte Carlo simulation, unbiased
estimates, numerical integration, finance, stochastic volatility model

\end{abstract}

\newpage

\section{Introduction.}

Suppose $X_{n},$ $n=0,1,2,...$ is a stochastic process such that
$E(X_{n})\rightarrow x_{\infty}$ as $n\rightarrow\infty.$ \ Typically
$X_{0}=x_{0}$ is a deterministic seed or arbitrary value initiating the
iteration and we are interested in the limiting value $x_{\infty}.$ The
sequence $\{X_{n}\}$ may be deterministic, obtained by using a numerical
integration scheme to approximate an integral, or a Newton-Raphson scheme to
approximate the root of an equation. It may be a ratio estimator estimating a
population ratio or the result of a stochastic or deterministic approximation
to a root or maximum. In general we will only assume that it is possible to
sample from the distribution of the stochastic process for a finite period,
i.e. sample $X_{1},X_{2},...X_{m}$ \ for fixed $m$. \ 

A common argument advanced in favour of the use Monte Carlo (MC) methods as an
alternative to numerical ones is that the MC estimator is usually unbiased
with estimable variance. By increasing the sample size we are assured by
unbiasedness that the estimator is consistent and we can produce, for any
sample size, a standard error of the estimator. \ The statistical argument is
advanced against the use of numerical methods that they do not offer easily
obtained estimates of error. The purpose of this brief note is to show that
this argument is flawed; generally any consistent sequence of estimators can
be easily rendered unbiased and an error estimate is easily achieved. We do
not attempt to merely reduce the bias, but by introducing randomization into
the sequence, to completely eliminate it. The price we pay is an additional
randomization inserted into the sequence and a possible increase in the mean
squared error (MSE).

\section{The debiased sequence and its variance}

Suppose $N$ is a random variable, independent of the sequence $\{X_{n},$
$n=0,1,2,...\}$ taking finite \textit{non-negative} integer values. Suppose
$Q_{n}=P(N\geq n)>0$ $\ $for all $n=1,2,...$ For a given sequence $X_{n},$
$n=0,1,2,...$ we define the first backward difference as $\nabla X_{n}%
=X_{n}-X_{n-1}.$ Define the random variable
\begin{align}
Y &  =x_{0}+\sum_{n=1}^{N}\frac{\nabla X_{n}}{Q_{n}}\nonumber\\
&  =x_{0}+\sum_{n=1}^{\infty}\nabla X_{n}\frac{I(n\leq N)}{Q_{n}}\nonumber
\end{align}
This can be written in the more general form
\begin{equation}
Y=X_{0}+\sum_{n=1}^{\infty}\nabla X_{n}F_{n}\text{ }\label{GF}%
\end{equation}
\ where $F_{n},n=1,2,...$are random variables with $E[F_{n}|X_{n},X_{n-1}]=1$
and \ for some value of $N<\infty$ we have $F_{i}=0$ for $i>N.$ \ We will show
that $Y$ is an unbiased estimator of the limit $x_{\infty}$ with easily
estimated standard error. \ It is obviously unbiased provided that it is
integrable and one can interchange the sum and the expected value since with
$\mathcal{F}=\sigma(X_{1},X_{2},.....),$ $E\left(  x_{0}+\sum_{n=1}^{\infty
}\nabla X_{n}\frac{I(n\leq N)}{Q_{n}}\right)  =E\left[  x_{0}+\sum
_{n=1}^{\infty}\nabla X_{n}E\left(  \frac{I(n\leq N)}{Q_{n}}|\mathcal{F}%
\right)  \right]  =x_{0}+\sum_{n=1}^{\infty}\nabla X_{n}=x_{\infty}.$

Assume for the calculation of the variance that $E(X_{n}-x_{\infty}%
)^{2}\rightarrow0$ as $n\rightarrow\infty.$ Then
\begin{align}
\sigma_{Y}^{2} &  =var(Y)=E[var(Y|\mathcal{F})]+var(E(Y|\mathcal{F}%
))\nonumber\\
&  =E[var(Y|\mathcal{F})]+var(x_{\infty})\nonumber\\
&  =E\left(  \sum_{n=1}^{\infty}\frac{(\nabla X_{n})^{2}}{Q_{n}^{2}}%
Q_{n}(1-Q_{n})+2\sum_{n=2}^{\infty}\sum_{j=1}^{n-1}\frac{\nabla X_{n}\nabla
X_{j}}{Q_{n}Q_{j}}Q_{n}(1-Q_{j})\right)  \nonumber\\
&  =E\left(  \sum_{n=1}^{\infty}\frac{(\nabla X_{n})^{2}}{Q_{n}}%
(1-Q_{n})+2\sum_{j=1}^{\infty}\sum_{n=j+1}^{\infty}\frac{\nabla X_{n}\nabla
X_{j}}{Q_{j}}(1-Q_{j})\right)  \label{varY}\\
&  =E\left(  \sum_{n=1}^{\infty}\frac{(\nabla X_{n})^{2}}{Q_{n}}%
(1-Q_{n})+2\sum_{j=1}^{\infty}\frac{(x_{\infty}-X_{j})\nabla X_{j}}{Q_{j}%
}(1-Q_{j})\right)  \nonumber\\
&  =E\left(  \sum_{n=1}^{\infty}\left[  (\nabla X_{n})^{2}+2(x_{\infty}%
-X_{n})\nabla X_{n}\right]  (\frac{1-Q_{n}}{Q_{n}})\right)  \nonumber\\
&  =\sum_{n=1}^{\infty}E\left(  \frac{\nabla X_{n}\left[  2x_{\infty}%
-X_{n}-X_{n-1}\right]  }{Q_{n}}-\sum_{n=1}^{\infty}\nabla X_{n}\left[
2x_{\infty}-X_{n}-X_{n-1}\right]  \right)  \nonumber\\
&  =\sum_{n=1}^{\infty}E\left(  \frac{2x_{\infty}\nabla X_{n}-\nabla X_{n}%
^{2}}{Q_{n}}-\sum_{n=1}^{\infty}\left(  2x_{\infty}\nabla X_{n}-\nabla
X_{n}^{2}\right)  \right)  \nonumber\\
&  =\sum_{n=1}^{\infty}E\left(  \frac{2x_{\infty}\nabla X_{n}-\nabla X_{n}%
^{2}}{Q_{n}}\right)  -\left(  2x_{\infty}(x_{\infty}-x_{0})-\left(  x_{\infty
}^{2}-x_{0}^{2}\right)  \right)  \nonumber\\
&  =\sum_{n=1}^{\infty}E\left(  \frac{2x_{\infty}\nabla X_{n}-\nabla X_{n}%
^{2}}{Q_{n}}\right)  -(x_{\infty}-x_{0})^{2}\nonumber\\
&  =\sum_{n=1}^{\infty}\frac{2x_{\infty}\nabla\mu_{n}-\nabla(\sigma_{n}%
^{2}+\mu_{n}^{2})}{Q_{n}}-(x_{\infty}-x_{0})^{2}\nonumber\\
&  =\sum_{n=1}^{\infty}\frac{2\left(  x_{\infty}-\xi_{j}\right)  \nabla\mu
_{j}-\nabla\sigma_{j}^{2}}{Q_{n}}-(x_{\infty}-x_{0})^{2}%
\end{align}
where $E[\nabla X_{n}^{2}]=EX_{n}^{2}-EX_{n-1}^{2}=\sigma_{n}^{2}+\mu_{n}%
^{2}-\sigma_{n-1}^{2}-\mu_{n-1}^{2}=\nabla(\sigma_{n}^{2}+\mu_{n}^{2})$ and
$\xi_{n}=\frac{\mu_{n}+\mu_{n-1}}{2}.$ Then $\sigma_{Y}^{2},$ as given in
(\ref{varY}) can be unbiasedly estimated using
\[
\widehat{\sigma}_{Y}^{2}=\sum_{n=1}^{N}\frac{(\nabla X_{n})^{2}}{Q_{n}^{2}%
}(1-Q_{n})+2\sum_{j=1}^{N}\sum_{n=j+1}^{N}\frac{\nabla X_{n}\nabla X_{j}%
}{Q_{j}^{2}Q_{n}}(1-Q_{j}).
\]

Suppose $N\geq n_{s}$ with probability one. Then the average over a large
number of values of $Y$, \ i.e. a large number of values of $N,\{X_{i}\}$
takes the form
\begin{align*}
\overline{Y} &  =\overline{X_{0}}+\sum_{n=1}^{N_{\max}}\overline{\nabla X_{n}%
}\frac{freq(n\leq N)}{Q_{n}},\\
&  =\overline{X_{n_{s}}}+\sum_{n=n_{s}+1}^{N_{\max}}\overline{\nabla X_{n}%
}\frac{freq(n\leq N)}{Q_{n}}%
\end{align*}
where $N_{\max}$ is the largest observed value of $N$ and $\overline{\nabla
X_{n}}$ \ denotes the average of the observed values of $\nabla X_{n}$ for
which the corresponding $N\geq n.$ This takes \ the form of (\ref{GF}) with
term $F_{n}=\frac{freq(n\leq N)}{Q_{n}}$ obtained from simulating values of
$N$ alone. \ Suppose we wish to minimize the variance subject to a constraint
on the expected value of $N,$ i.e.
\[
\min_{Q}\{\sum_{n=1}^{\infty}\frac{2x_{\infty}\nabla\mu_{n}-\nabla(\sigma
_{n}^{2}+\mu_{n}^{2})}{Q_{n}}-(x_{\infty}-x_{0})^{2}\}
\]%
\[
\text{subject to }\sum_{n}Q_{n}=\mu.\text{ \ }%
\]
Of course we also require that $Q_{n}$ is non-increasing and positive but for
the moment we will ignore these additional constraints. Then we obtain, on
differentiating the Lagrangian with respect to $Q_{n},$with $\xi_{n}=\frac
{\mu_{n}+\mu_{n-1}}{2},$
\begin{equation}
Q_{n}\sim c\sqrt{|2\left(  x_{\infty}-\xi_{n}\right)  \nabla\mu_{n}%
-\nabla\sigma_{n}^{2}|}\label{optimalQ}%
\end{equation}
where $c$\ is determined by the constraint $\sum_{n}Q_{n}=\mu$ and the minimum
variance is finite provided that
\[
\sum_{j=1}^{\infty}\sqrt{|2\left(  x_{\infty}-\xi_{j}\right)  \nabla\mu
_{j}-\nabla\sigma_{j}^{2}|}<\infty.
\]
\ While this is not entirely practical because it requires $x_{\infty},$ it is
common to have some information on the rate of convergence of the sequence
that can be used to design an asymptotically appropriate sequence $Q_{n}.$ For
example if we believe $x_{\infty}-X_{n}\sim ar^{n}$ \ for some $r<1$ and $a,$
then we might choose $Q_{n}\sim cr^{n/2}$ or a random variable $N$ which has a
geometric distribution, at least in the tails. Suppose the sequence $X_{n}$ is
deterministic and we use $Q_{n}\sim c\sqrt{|\nabla x_{n}|}$. \ Then the
variance is finite provided the series $\sum_{n}\sqrt{\nabla x_{n}}$is convergent.

\begin{figure}
[ptb]
\begin{center}
\includegraphics[
height=2.7129in,
width=5.4544in
]%
{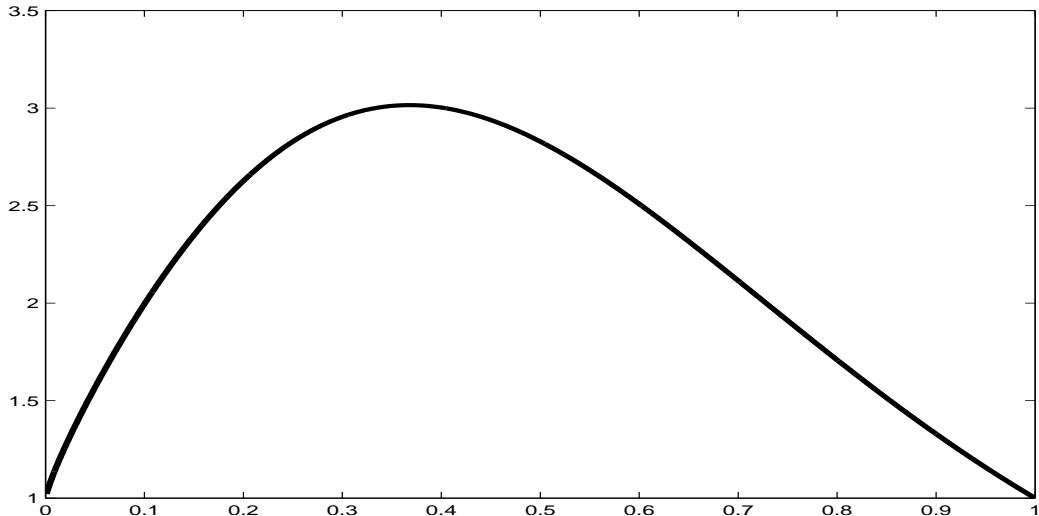}%
\caption{Relative increase in MSE due to debiasing the sequence.}%
\label{plotr}%
\end{center}
\end{figure}

Let us consider a simple example before we look at more complex ones.
\ Suppose $X_{n}=b+ar^{n}$ \ for $n=1,...,|r|<1$ and $x_{\infty}=b.$ \ Then
$\nabla X_{n}=ar^{n-1}(r-1),n\geq2$ and $\nabla X_{n}^{2}=2abr^{n-1}%
(r-1)+a^{2}r^{2n-2}(r^{2}-1)\ $and, as we already verified more generally,
$E(X)=E\left[  X_{0}+\sum_{n=1}^{\infty}\nabla X_{n}\frac{I(n\leq N)}{Q_{n}%
}\right]  =b$ \ whatever the distribution of $N.$ \ Suppose we use a shifted
geometric distribution for $N$ so that $P(N\geq n)=q^{n-s},n=s,s+1,...$ for
$|q|<1$. \ Evidently to minimize the variance we should choose
\begin{equation}
Q_{n}\sim c|r|^{n},n=1,...
\end{equation}
so that $q=|r|.$ The variance for general $q$ is%

\begin{align*}
\sum_{n=s+1}^{\infty}\frac{2\left(  x_{\infty}-\xi_{n}\right)  \nabla X_{n}%
}{Q_{n}}-(x_{\infty}-X_{s})^{2}  &  =\sum_{n=s+1}^{\infty}\frac{-2a\left(
\frac{r^{n}+r^{n-1}}{2}\right)  ar^{n-1}(r-1)}{q^{n-s}}-a^{2}r^{2s}\\
&  =a^{2}r^{2s}\left[  \frac{1-q}{q-r^{2}}\right]  \text{ \ where }1>q>r^{2}.
\end{align*}
Suppose we wish to minimize this over the values of $s$ and $q$ subject to the
constraint that $E(N)=\frac{q}{1-q}+s=\mu_{N}$ is constant (ignoring the
integer constraint on $s$). Then with $z=\frac{1}{q},$%

\[
\min_{s,z}r^{2s}\left[  \frac{z-1}{1-r^{2}z}\right]  \text{ subject to }%
\frac{1}{z-1}+s=\mu_{N}\text{ or }%
\]%
\[
\min_{z}r^{-2/(z-1)}\left[  \frac{z-1}{1-r^{2}z}\right]  ,\text{ for }\frac
{1}{r^{2}}>z>1
\]
which minimum occurs when $z=\frac{1}{r},$ or $q=|r|$ and $s\simeq\mu
_{N}-\frac{|r|}{1-|r|}$ and then the minimum variance is
\[
a^{2}|r|^{2s-1}\simeq a^{2}|r|^{2\mu_{N}-\frac{2|r|}{1-|r|}}%
\]
Notice that the mean squared error, if we were to stop after $\mu_{N}$
iterations,\ is $a^{2}|r|^{2\mu_{N}}$ so we have purchased unbiasedness of the
estimator at a cost of increasing the MSE by a factor of approximately
$|r|^{-\frac{2|r|}{1-|r|}}.$ \ This factor is plotted in Figure \ref{plotr}.
\ It can be interpreted as follows:\ in the worst case scenario when $p$ is
around .4, \ we will need about 3 times the sample size for the debiased
estimator to achieve the same mean squared error as a conventional iteration
using determinisitic $N.$ However when $|r|$ is close to $1$ indicating a
slower rate of convergence, there is very little relative increase in the MSE.%

%TCIMACRO{\FRAME{ftbpFU}{5.4544in}{2.7129in}{0pt}{\Qcb{Relative increase in MSE
%due to debiasing the sequence.}}{\Qlb{plotr}}{plotr.eps}%
%{\special{ language "Scientific Word";  type "GRAPHIC";  display "USEDEF";
%valid_file "F";  width 5.4544in;  height 2.7129in;  depth 0pt;
%original-width 6.5259in;  original-height 5.188in;  cropleft "0";
%filename 'plotr.eps';file-properties "XNPEU";}} }%
%BeginExpansion
%EndExpansion

Note: \ the optimisation problem above tacitly assumed that the computation
time required to generate the sequence is $O(n).$ \ This is not the case with
some applications; for example in the numerical integral below the computation
time is $O(2^{n})$ since there are $2^{n}$ intervals and $2^{n}+1$ function
evaluations and in this case \ a more appropriate minimization problem, having
budget constraint $E(2^{N})=2^{s}\frac{1-q}{1-2q}=c\geq2^{s},$ is, with
$0<q<\frac{1}{2},$
\[
\min_{s,q}r^{2s}\left[  \frac{1-q}{q-r^{2}}\right]  \text{ \ subject to }%
\frac{1}{2}>q>r^{2}\text{ and \ }2^{s}\frac{1-q}{1-2q}=c>2^{s},s=0,1,2,...
\]
or, putting $q=\frac{c-2^{s}}{2c-2^{s}},$ \
\[
\min_{s\leq\log_{2}(c)}r^{2s}\left[  \frac{c}{c-2^{s}-r^{2}\left(
2c-2^{s}\right)  }\right]
\]
which minimum appears to occur for $s=\left[  \log_{2}(c)\right]  $ \ \ and
$q=\frac{c-2^{s}}{2c-2^{s}}$ when $r<0.53$ and otherwise $s$ may be somewhat
smaller. Intuitively, when the rate of convergence is reasonably fast ( so $r$
is small) \ then the minimum variance is achieved by a large guarantee on the
value of $N$ ($s$ large) and then the residual budget $(c-2^{s})$ used to
produce unbiasedness by appropriate choice of $q.$

\section{Examples}

\begin{example}
Unbiased Estimation of a root
\end{example}

Suppose we wish to find the root of a nonlinear function $h$. For a toy
example, suppose we wish to solve for $x$ the equation
\[
h(x)-\alpha=0.
\]
We might wish to use (modified) Newton's method with a random starting value
to solve this problem,\ requiring randomly generating the initial value
$X_{0}$ and then iterating
\[
X_{n+1}=X_{n}-\delta_{n}\text{ \ where }\delta_{n}=\min(1,\max(-1,\frac
{h(X_{n})-\alpha}{h^{\prime}(X_{n})}))
\]
but of course after a finite number of steps, the current estimate $X_{n}$ is
likely a \ biased estimator of the true value of the root. \ We implemented
the debiasing procedure above with $h(x)=x^{3}$ and $\alpha=1$. We generated
$X_{0}$ from a $U(-2,3)$ distribution, chose $P(N=n)=(1-p)p^{n-s}%
,n=s,s+1,...$and for simplicity used $s=4,p=\frac{3}{4}$ and repeated for
$100,000$ simulations. \ The sample mean of the estimates was $1.0023$ \ and
the sample variance $0.011$. \ Although the procedure works well in this case
when we start sufficiently close to the root, it should be noted that this
example argues for an adaptive choice of $Q_{n},$ one which permits a larger
number of iterations (larger values of $N)$ when the sequence seems to
indicate that we have not yet converged. This is discussed below.\bigskip

\noindent\textbf{Stopping times for }$N.$

In view of the last example, particularly if $X_{0}$ is far from $x_{\infty},$
it would appear desirable to allow $N$ to be a stopping time. In order to
retain unbiasedness, it is sufficient that%

\begin{equation}
EY=E\left[  x_{0}+\sum_{n=1}^{\infty}\nabla X_{n}\frac{I(n\leq N)}{Q_{n}%
}\right]  =E\left[  x_{0}+\sum_{n=1}^{\infty}\nabla X_{n}\right]  \nonumber
\end{equation}
or
\[
E\left[  \frac{I(n\leq N)}{Q_{n}}|X_{1},X_{2},...X_{n}\right]  =1.
\]
Therefore it is sufficient that $Q_{n}=P(N\geq n|X_{1},X_{2},...X_{n})$ and
one simple rule for an adaptive construction of $N$ is: \
\[
P(N\geq n|X_{1},X_{2},...X_{n})=\left\{
\begin{array}
[c]{ccc}%
P(N\geq n-1|X_{1},X_{2},...X_{n-1}) & \text{if} & \nabla X_{n}>\varepsilon\\
pP(N\geq n-1|X_{1},X_{2},...X_{n-1}) & \text{if} & \nabla X_{n}\leq\varepsilon
\end{array}
\right.
\]
There are, of course, many potential more powerful rules for determining the
shift in the distribution of $N$ \ but we we concentrate here on establishing
the properties of the simplest version of this procedure.

\begin{example}
Simpson's rule. \ 
\end{example}

Consider using a trapezoidal rule for estimating the integral
\[
I_{\infty}=\int_{0}^{1}f(x)dx
\]
using $2^{n}+1$ function evaluations which evaluate the function on the grid
$0,\Delta x,2x,...2^{n}\Delta x=1$. \ Denote the estimate of the integral
$I_{n}.$ Here $\Delta x=2^{-n}$ \ and the error in Simpson's rule assuming
that the function has bounded fourth derivative is $O((\Delta x)^{4}%
)=O(\left(  2^{-n}\right)  ^{4})=O(16^{-n})$. This suggests a random $N$ such
that $Q_{n}\sim\frac{1}{4^{n}}$ or a (possibly shifted) geometric distribution
with $p=\frac{3}{4}.$ Suppose $Q_{n}=4^{-n+2},n=2,3,...$ This means
$E(N)=\frac{7}{3}$ \ which is quite small. In general, the estimator has
finite variance since
\[
\sum_{n=1}^{\infty}\frac{\nabla I_{n}}{Q_{n}}<\infty.
\]
More generally, if $N$ \ has a shifted geometric distribution with probability
function $P(N=n)=p(1-p)^{n-s},n=s,s+1,...$ parameter $p,$ the expected number
of function evaluations in the quadrature rule is
\begin{align*}
\sum_{n=1}^{\infty}(2^{n}+1)P(N  &  =n)=\sum_{n=s}^{\infty}(2^{n}%
+1)(1-p)^{n-s}p\\
&  =1+2^{s}\frac{p}{2p-1}%
\end{align*}
and this is $7,$ for example, when $p=\frac{3}{4},s=2.$ How well does this
perform? This provides an unbiased estimator of the integral with variance
\[
\sigma_{X}^{2}=\sum_{n=2}^{\infty}\frac{\nabla I_{n}(1-Q_{n})}{Q_{n}}\left[
2I_{\infty}-I_{n}-I_{n-1}\right]
\]
which can be evaluated or estimated in particular examples and compared with
the variance of the corresponding crude Monte Carlo estimator. For a
reasonable comparison, the latter should have the same (expected) number of
function evaluations, i.e. 7 and therefore has variance
\[
\frac{1}{7}\left(  \int_{0}^{1}f^{2}(x)dx-I_{\infty}^{2}\right)
\]
Take, for example, the function $f(x)=\sin(\pi x)$ so that \ $I_{\infty}%
=\int_{0}^{1}\sin(\pi x)dx=\frac{2}{\pi}=0.63662$ and $\int_{0}^{1}(\sin(\pi
x))^{2}dx=\frac{1}{2}.$ \ In this case the variance of the MC estimator with
seven function evaluations is $\frac{1}{7}\left(  0.5-\left(  \frac{2}{\pi
}\right)  ^{2}\right)  \simeq0.013531.$ We compare this with the estimator
obtained by randomizing the number of points in a Simpson's rule. Here it is
easy to check that

\begin{center}%
\begin{tabular}
[c]{||l|l|l|l|l|l|l||}\hline\hline
$n$ & {\small 1} & {\small 2} & {\small 3} & {\small 4} & {\small 5} &
{\small 6}\\\hline
$I_{n}$ & {\small 0.3047} & {\small 0.2149} & {\small 0.2124} &
{\small 0.2122} & {\small 0.2122} & {\small 0.2122}\\\hline
$\nabla I_{n}$ & {\small -0.089821} & {\small -0.002562} & {\small -0.000139}
& {\small -0.000008} & {\small -0.000001} & {\small -0.00000003}\\\hline\hline
\end{tabular}

Table 1: Values of the numerical integral $I_{n}$\ and $\nabla I_{n}$ with
$2^{n}$ intervals

\bigskip
\end{center}

\noindent and in this case the variance of the debiased Simpson's rule
estimate is $\sigma_{X}^{2}\simeq6.41\times10^{-6}$ indicating more than a two
thousand-fold gain in efficiency over crude Monte Carlo.

\noindent\textbf{Note:} We have chosen the grid size $2^{-n}$ in view of the
fact that when $N=n,$ we need the integrals \thinspace$I_{j}$ for all $j\leq
n.$ In this case, we can simply augment the function evaluations we used for
$I_{j}$ in order to obtain $I_{j+1}.$

The major advantage of this debiasing procedure however is not as a
replacement for Crude Monte Carlo in cases where unbiased estimators exist,
but as a device for creating unbiased estimators when their construction is
not at all obvious. This is the case whenever the function of interest is a
nonlinear function of variables that can be easily and unbiasedly estimated as
in the following example.

\begin{example}
Heston Stochastic Volatility model
\end{example}

In the Heston stochastic volatility model, under the risk neutral measure $Q,
$ the price of an asset $S_{t}$ and the volatility process $V_{t}$ are
governed by the pair of stochastic differential equations%

\begin{align*}
dS_{t}  &  =rS_{t}dt+\sqrt{V_{t}}S_{t}\rho dW_{1}(t)+\sqrt{1-\rho^{2}}%
dW_{2}(t),\text{ \ \ }S_{0}=s_{0}\\
dV_{t}  &  =\kappa(\theta-V_{t})dt+\sigma_{V}\sqrt{V_{t}}dW_{1}(t),\text{
\ \ \ \ \ \ \ \ \ \ \ \ \ \ \ \ \ \ \ \ }V_{0}=v_{0}%
\end{align*}
where $W_{1}(t)$ and $W_{2}(t)$ are independent \ Brownian motion processes,
\ $r$ is the interest rate, $\rho$ is the correlation between the Brownian
motions driving the asset price \ and the volatility process, $\theta$ is the
long-run level of volatility and $\kappa$ is a parameter governing the
\ Denote by $BS(S_{0},K,r,T,\sigma)$ \ the Black-Scholes price of a call
option having initial stock value $S_{0},$ volatility $\sigma$, interest rate
$r,$ expiration time $T,$ option stike price $K$ and $0$ \ dividend yield.
\ The price of a call option in the Heston model can be written as an expected
value under the risk-neutral measure $Q$ of a function of two variables
$g(V_{T},I(T))$ (see for example Willard (1997) and Broadie and Kaya (2006))%
\begin{align*}
&  e^{-rT}E^{Q}(S_{T}-K)^{+}\text{ \ }=E[BS(S_{0}\xi,K,r,T,\widetilde{\sigma
}\sqrt{1-\rho^{2}})]\text{ \ where }\\
\xi &  =\xi(V_{T},I(T))=\exp(-\frac{\rho^{2}}{2}I(T)+\rho\int_{0}^{T}%
\sqrt{V_{s}}dW_{1}(s))\\
&  =\exp(-\frac{\rho^{2}}{2}I(T)+\frac{\rho}{\sigma}(V_{T}-V_{0}+\kappa
I(T)-\kappa\theta T))\text{ and}\\
\widetilde{\sigma}  &  =\widetilde{\sigma}(I(T))=\sqrt{\frac{I(T)}{T}}\text{
\ where }I(T)=\int_{0}^{T}V_{s}ds.
\end{align*}
This can be valued conditionally on $V_{T},I(T)$ with the usual Black-Scholes
formula. In particular with $g(V_{T},I(T))=BS(S_{0}\xi,K,r,T,\widetilde
{\sigma}\sqrt{1-\rho^{2}}),$ \ the option price is $E^{Q}g(V_{T},I(T)).$

Note that $g$ is clearly a highly nonlinear function of $V_{T}$ and $I(T)$
\ and so, even if exact simulations of the latter were available, it is not
clear how to obtain an unbiased simulation of $g.$ In the Heston model, and
indeed various other stochastic volatility models, it is possible to obtain an
exact simulation of the value of the process $V_{t}$ at finitely many values
of $t$ \ so it is possible to approximate the integral $I(T)$ using $I_{n}(T)$
\ obtained from a trapezoidal rule with $1+2^{n}$ points. This raises the
question of what we should choose as a distribution for $N.$ Under conditions
on the continuity of the functional of the process whose expected value is
sought, \ Kloeden and Platen (1995, Theorem 14.1.5, page 460) show that the
Euler approximation to the process with interval size $2^{-n}$ results in an
error in the expected value of order $2^{-n\chi}$ \ where $\chi=1$ \ for
sufficiently smooth (four times continuously differentiable) drift and
diffusion coefficients so for simplicity consider this case. \ This implies
that
\[
|Eg(V_{T},I_{n}(T))-Eg(V_{T},I(T))|<\text{constant }\times2^{-n}.
\]
which suggests we choose $Q_{n}\sim2^{-n}$.

As before we randomly generate $N$ from a (possibly shifted) geometric$(p)$
distribution with $p=\frac{1}{2}$. The function to be integrated $V_{s}$ is
not twice differentiable so we need to determine empirically the amount of the
shift (and we experimented with reasonable values of $p$). \ We chose
parameters $p=0.5$ and shifted the geometric random variable by $s=2$ so that
$P(N=n)=p(1-p)^{n-s}$ for $n=s,s+1,....$ The parameters used in our simulation
were taken from Broadie and Kaya(2004): $s_{0}=100;K=100;V_{0}=0.09;\kappa
=2.00;\theta=0.09;r=.05;\sigma=1;\rho=-.3;T=5;$ \ for which, according to
Broadie and Kaya, the true option price \ is around 34.9998. \ 1,000,000
\ simulations with $p=0.75$ and $s=4$ provided an estimate of this option
price of 34.9846 \ with a standard error \ of 0.0107 so there is no evidence
of bias in these simulations. \ With parameter values $\theta=0.019;\kappa
=6.21;\sigma=0.61;v_{0}=0.010201;r=0.0319;$ $s_{0}=100;K=100;$ $T=1;\rho=-0.7$
and $p=0.75$ and $s=4,$ we conducted $10^{6}$ simulations leading to an
estimate of 6.8115 with a standard error of 0.0048998. This is in agreement
with the Broadie and Kaya "true option price" of 6.801. Note that the Feller
condition for positivity requires $2\kappa\theta>\sigma^{2}$ \ which fails in
the above cases. This means that the volatility process hits zero with
probability one, and for some parameter values, it does so frequently which
may call into question the value of this model with these parameter values.
100,000 simulations from these models used about 10-13 minutes running Matlab
5.0 on an intel Core 2 Quad CPU @2.5 GHz.

\section{Conclusion}

When numerical methods such as quadrature or numerical solutions to equations
may result in a biased estimator, a procedure is suggested which eliminates
this bias and provides statistical estimates of error. This procedure is
successfully implemented both in simple root finding problems and in more
complicated problems in finance and has enormous potential for providing Monte
Carlo extensions of numerical procedures which allow unbiased estimates and
error estimates. \


\bigskip

\begin{thebibliography}{9}                                                                                                %


\bibitem {}Willard, G. A. Calculating prices and sensitivities for path
independent derivative securities in multifactor models. \textit{J.Derivatives
}5(1) 45--61 \ (1997)

\bibitem {}Broadie, M. and Kaya, O. \ Exact Simulation of Stochastic
Volatility and Other Affine Jump Diffusion Processes \textit{Operations
Research} 54 (2), 217--231 (2006)

\bibitem {}Glasserman, P.: \ \textit{Monte Carlo Methods in Financial
Engineering.} Springer-Verlag, New York (2003)

\bibitem {}Kloeden, P.E. and Platen, E.: \textit{Numerical Solution of
Stochastic Differential Equations (Stochastic Modelling and Applied
Probability), }Springer, New York \ (1995)
\end{thebibliography}
\end{document}